\documentclass[authoryear,preprint,review,12pt]{elsarticle}

\usepackage{graphicx}

\usepackage{amssymb}

\journal{Physics Letters B}

\begin{document}

\begin{frontmatter}



\title{Superdeformation in Asymmetric N$>$Z Nucleus $^{40}$Ar}


\author[CNS]{E.~Ideguchi\corref{cor1}}
\cortext[cor1]{Corresponding author}
\ead{ideguchi@cns.s.u-tokyo.ac.jp}
\author[CNS]{S.~Ota}
\author[Kyushu]{T.~Morikawa}
\author[JAEA]{M.~Oshima}
\author[JAEA]{M.~Koizumi}
\author[JAEA]{Y.~Toh}
\author[JAEA]{A.~Kimura}
\author[JAEA]{H.~Harada}
\author[JAEA]{K.~Furutaka}
\author[JAEA]{S.~Nakamura}
\author[JAEA]{F.~Kitatani}
\author[JAEA]{Y.~Hatsukawa}
\author[JAEA]{T.~Shizuma}
\author[CIT]{M.~Sugawara}
\author[KEK]{H.~Miyatake}
\author[KEK]{Y.X.~Watanabe}
\author[KEK]{Y.~Hirayama}
\author[Senshu]{M.~Oi}

\address[CNS]{Center for Nuclear Study, The University of Tokyo, Wako, Saitama 351-0198, Japan}
\address[Kyushu]{Department of Physics, Kyushu University, Hakozaki, Fukuoka 812-8581, Japan}
\address[JAEA]{Japan Atomic Energy Agency, Tokai, Ibaraki 319-1195, Japan}
\address[CIT]{Chiba Institute of Technology, Narashino, Chiba 275-0023, Japan}
\address[KEK]{Institute of Particle and Nuclear Studies, 
High Energy Accelerator Research Organization (KEK), 1-1 Oho, Tsukuba, Ibaraki 305-0801, Japan}
\address[Senshu]{Institute of Natural Sciences, Senshu University, Tokyo 101-8425, Japan}

\begin{abstract}
A rotational band with five $\gamma$-ray transitions ranging from 
2$^{+}$ to 12$^{+}$ states was identified in $^{40}$Ar.
This band is linked through 
$\gamma$ transitions from the excited 2$^{+}$, 4$^{+}$ 
and 6$^{+}$ levels to the low-lying states; this determines the excitation energy 
and the spin-parity of the band.
The deduced transition quadrupole moment of 1.45$^{+0.49}_{-0.31} eb$ indicates
that the band has a superdeformed shape.
The nature of the band is revealed by cranked Hartree--Fock--Bogoliubov 
calculations and a multiparticle--multihole configuration is assigned to the band. 
\end{abstract}


\end{frontmatter}





Until recently, the nuclear magic numbers have been considered to be robust 
and durable with respect to the shell structure of atomic nuclei.
The nuclear shell model initiated by Mayer and Jensen \cite{MayJen}
successfully accounts for the shell structure of nuclei on and near 
the $\beta$-stability line. 
However, recent experimental studies on neutron-rich nuclei 
far from the $\beta$-stability line have revealed 
disappearance of magic numbers 
(e.g., N $=$ 8 \cite{be12a,be12b,be12c,be12d}, 20 \cite{mg32}) 
and appearance of new magic numbers (e.g., N $=$ 16 \cite{n16magic}).
The new N $=$ 16 magic number comes from the reduction of the N $=$ 20
shell gap. 
One theoretical explanation \cite{otsuka1, utsuno1} for this phenomenon 
is that the attractive 
monopole part of the tensor force acting between the $\pi$d$_{5/2}$ and 
$\nu$d$_{3/2}$ orbitals 
reduces the N $=$ 20 gap in nuclei with high N/Z ratios. 

Another anomalous phenomenon discovered in neutron-rich nuclei is 
the occurrence of a {\it strongly deformed} ground state on and near 
the N $=$ 20 magic nuclei in the Z $\sim$ 12 region, 
the so-called 'island of inversion' \cite{inversion}. 
Here, the intruder two-particle, two-hole (2p--2h) configuration occupying 
the {\it fp} shell and the normal 0p--0h configuration in the {\it sd} shell 
are inverted in energy, and the 2p--2h configuration dominates in the 
ground state. 
This is a new challenge to the nuclear shell model. 
The most convincing evidence for the large quadrupole collectivity of the nuclei 
in the 'island of inversion' was the measured low E(2$_{1}^{+}$) energy \cite{mg32a} 
and the high B(E2) strength \cite{mg32} as well as the enhancement of the binding 
energy \cite{mg32mass}. 
The experimentally deduced large B(E2; 0$_{g.s.}^{+}\rightarrow$2$_{1}^{+}$) value 
indicates large deformation ($\beta_{2}=$ 0.512(44)) in $^{32}$Mg \cite{mg32}.
$^{34}$Mg (N $=$ 22) shows an even larger deformation ($\beta_{2}=$ 0.58(6)) 
\cite{mg34}.
Monte Carlo shell model calculations \cite{mcsm} that include the effects of 
the narrowed shell gap at N $=$ 20 and enhanced 2p--2h excitations reproduce 
the experimental values quite well \cite{mg34}. 

On the other hand, such a multiparticle--multihole (mp--mh) excitation appears 
in the excited states of nuclei near the $\beta$-stability line.
These states can be studied by the heavy-ion induced fusion-evaporation reaction
of stable isotopes. 
In fact, an mp--mh excitation from the {\it sd} to {\it fp} shell produces 
superdeformation (SD) in N $=$ Z light mass nuclei, 
such as $^{36}$Ar \cite{ar36}, $^{40}$Ca \cite{ca40}, and $^{44}$Ti \cite{ti44}.
These SD nuclei exhibit a large deformation of $\beta_{2}\sim$ 0.5, which is about 
the same magnitude as the ground state deformation in the 'island of inversion'.
The presence of SD in the three abovementioned nuclei can be also understood 
in terms of the SD shell gaps of N $=$ Z $=$ 18, 20, and 22, respectively \cite{ca40}. 
In this region, the spherical and SD magic numbers occur at similar 
particle numbers, which results in shape coexistence. 
However, the existence of a SD shell structure in neutron-rich nuclei 
has not been experimentally confirmed yet.

In order to access the currently reachable neutron-richest SD states with asymmetric 
SD magic numbers, especially the nucleus with N $=$ 22 corresponding to $^{34}$Mg, 
we employed a proton emission channel (2p2n) in the fusion-evaporation reaction using 
the neutron-richest beam and target combination of stable isotopes obtained so far. 
Consequently, we successfully populated the high-spin states of the SD double magic
Z $=$ 18 and N $=$ 22 nucleus, $^{40}$Ar. 
In this Letter, we report experimental results on the SD states in $^{40}$Ar 
associated with the mp--mh excitation between the {\it sd} and {\it fp} shells.

High-spin states in $^{40}$Ar have previously been studied by proton-$\gamma$ 
coincidence measurements using the $^{37}$Cl($\alpha$,~p$\gamma$) reaction 
\cite{old40ar}.
High-spin levels below 6.8~MeV were identified up to (8$^{+}$) and spin-parity 
assignments up to the 6$^{+}$ state were obtained from the particle-$\gamma$ 
angular correlations. 
The parity of the 5$^{-}$ state at 4.494~MeV was determined by the linear 
polarization of the 5$^{-}\rightarrow$4$^{+}$ transition at 1602~keV.
The lifetimes of low-lying levels were measured by the Doppler-shift attenuation 
method. 
The high E2 strengths of the 6$_{2}^{+}\rightarrow$4$_{2}^{+}$ 
and 4$_{2}^{+}\rightarrow$2$_{2}^{+}$ transitions are respectively deduced to be 
67$_{-19}^{+38}$ and 46$_{-10}^{+15}$ in Weisskopf units, 
which indicates the large collectivity of the band. 
However, the (8$^{+}$) assignment was based solely on the similarity of the
level structure to that in $^{42}$Ca, but the $\gamma-\gamma$ coincidence
relations of the in-band transitions were not examined and the presence of the 
band structure was not unambiguously identified by experiment. 
Therefore, it is essential to find the higher-spin members of the rotational band 
and to confirm the coincidence relations between the in-band $\gamma$ transitions.

High-spin states in $^{40}$Ar were populated via the 
$^{26}$Mg($^{18}$O, 2p2n)$^{40}$Ar reaction with a 70-MeV $^{18}$O beam
provided by the tandem accelerator facility at the Japan Atomic Energy Agency.
Two stacked self-supporting foils of $^{26}$Mg enriched isotopes with 
thicknesses of 0.47 and 0.43 mg/cm$^{2}$ were used. 
The mean beam energy of the $^{18}$O beam used to irradiate the $^{26}$Mg foils 
was 69.0~MeV.
Gamma rays were measured by the GEMINI-II array \cite{gemini} consisting of 
16 HPGe detectors with BGO Compton suppression shields, in coincidence with 
charged particles detected by the Si-Ball \cite{siball}, a 4 $\pi$ array 
consisting of 11 $\Delta$E Si detectors that were 170~$\mu$m thick. 
The most forward placed Si detector was segmented into five sections and 
the other detectors were segmented into two sections each, giving a total of 
25 channels that were used to enhance the selectivity of multi charged-particle 
events.
With a trigger condition of more than two Compton-suppressed Ge detectors
firing in coincidence with charged particles, a total number of 
6.6$\times$10$^{8}$ events were collected.

Based on the number of hits in the charged particle detectors, events were 
sorted into three types of E$_{\gamma}-$E$_{\gamma}$ coincidence matrices 
for each evaporation channel.
A symmetrized matrix was created and the RADWARE program ESCL8R \cite{radware}
was used to examine the coincidence relations of $\gamma$ rays. 
By gating on the previously reported $\gamma$ rays, high-spin states
in $^{40}$Ar were investigated.

By gating on the known 1461, 1432, and 571~keV peaks of the 
2$^{+}\rightarrow $0$^{+}$, 
4$^{+}\rightarrow $2$^{+}$, and 
6$^{+}\rightarrow $4$^{+}$ transitions respectively, 
several new levels were identified above the 5$^{-}$ states at 4.49~MeV by 
connecting with high-energy $\gamma$ transitions of $\geq$2.5~MeV.
The previously assigned deformed band members of 2$_{2}^{+}$, 4$_{2}^{+}$, 
and 6$_{2}^{+}$ states were confirmed at 2.522, 3.515, and 4.960~MeV, 
respectively.
In addition, two $\gamma$-ray cascade transitions of 2269 and 2699~keV 
were identified in coincidence with the 993, 1445, and 1841~keV transitions, 
which form a rotational band up to the (12$^{+}$) state at 11.769~MeV. 
Linking $\gamma$ transitions were also observed between the excited 
2$_{2}^{+}$, 4$_{2}^{+}$, and 6$_{2}^{+}$ states and the low-lying 
2$_{1}^{+}$ and 4$_{1}^{+}$ levels, which establishes the excitation energies 
and the spin-parity assignment of the band.

Spins of the observed levels are assigned on the basis of the 
DCO (Directional Correlations from Oriented states) ratios of $\gamma$ rays 
by analyzing an asymmetric angular correlation matrix. 
The multipolarities of the in-band transitions of the band and 
the linking transitions of 4$_{2}^{+}\rightarrow $2$_{1}^{+}$ and 
6$_{2}^{+}\rightarrow $4$_{1}^{+}$ are consistent with a stretched quadrupole 
character. 
Assuming E2 multipolarity for the stretched quadrupole transition, the parity 
of the band was assigned to be positive.
The multipolarity of the 2699~keV transition could not be determined due to 
the lack of statistics, but it was in coincidence with other $\gamma$ 
transitions in the band and assigned as E2. 

To determine the deformation of the band, the transition quadrupole moment Q$_{t}$ 
was deduced. 
Lifetimes were estimated by the \cite{DSAM} technique, which is based on the 
residual Doppler shift of the $\gamma$-ray energies emitted from the deceleration 
of recoiling nuclei in a thin target. 
The average recoil velocity $<\beta>$ is expressed as a function of the 
initial recoil velocity to obtain $F(\tau) \equiv <\beta>/\beta_{0}$.
In Fig.~3, the fractional Doppler shift $F(\tau)$ is plotted against the 
$\gamma$-ray energy. 
The experimental $F(\tau)$ values are compared with the calculated values
based on known stopping powers \cite{srim2003}. 
In this calculation, the side feeding into each state is assumed to consist
of a cascade of two transitions having the same lifetime as the in-band
transitions.
The intensities of the side-feeding transitions were modeled to reproduce
the observed intensity profile.
The data are best fitted with a transition quadrupole moment 
$Q_{t} = 1.45_{-0.31}^{+0.49} e$b, which corresponds to a quadrupole 
deformation of $\beta_{2}=0.53_{-0.10}^{+0.15}$.
This result is consistent with a superdeformed shape of the band.

In order to compare the high-spin behavior of the rotational band in $^{40}$Ar
with the SD bands in $^{36}$Ar and $^{40}$Ca, 
the so-called 'backbending' plot of the SD bands is displayed in Fig.~\ref{fig4}.
The gradients of the plots correspond to the kinematic moments of inertia. 
Because $^{40}$Ar has a similar gradient to $^{36}$Ar and $^{40}$Ca, 
the deformation size of the $^{40}$Ar rotational band is expected to be as 
large as the deformation of the SD bands in $^{36}$Ar and $^{40}$Ca. 
Unlike $^{36}$Ar, no backbending was observed in $^{40}$Ar.
Its behavior is rather similar to that of $^{40}$Ca.
Many theoretical models, including the shell model \cite{ar36,caurier,poves,sun} 
and various mean-field models \cite{inakura,bender,HFB}, have been used to analyze 
$^{36}$Ar.
All the calculations reveal that the strong backbending in 
$^{36}$Ar is due to the simultaneous alignment of protons and neutrons 
in the f$_{7/2}$ orbital.

The pronounced difference in the high-spin behaviors of $^{36}$Ar and 
$^{40}$Ar implies that the addition of four neutrons to $^{36}$Ar gives rise to  
a dramatic effect on its structure. 
In order to understand this structural change theoretically, cranked 
Hartree--Fock--Bogoliubov (HFB) calculations with the P+QQ force \cite{HFB} 
were conducted.
The evolution of the nuclear shape was treated in a fully self-consistent manner 
and the model space of the full {\it sd-fp} shell plus the g$_{9/2}$ orbital 
was employed.
The calculation shows that $\beta_2 = 0.57$ at $J = 0 \hbar$ and that the deformation 
gradually decreases to 0.45 at $J = 12\hbar$. 
Triaxiality is found to be almost zero ($\gamma\simeq 0$) throughout this 
angular momentum range. 
This result agrees with the experimental $Q_t$ value within the error bars.

The occupation number of each orbital was also calculated (Table \ref{occupation}).
The ground-state configuration is expressed as 
({\it sd})$^{-2}$({\it fp})$^{2}$ relative to the ground-state configuration of 
$^{40}$Ca, where the Fermi levels for protons and neutrons lie at d$_{3/2}$
and f$_{7/2}$, respectively.
The self-consistently calculated second $0^+$ state has 
the ({\it sd})$^{-6}$({\it fp})$^{6}$ configuration.
Here, the {\it fp} shell is occupied by two protons and four neutrons,
while the {\it sd} shell has four proton holes and two neutron holes.
Considering the rise of the neutron Fermi level relative to that in $^{36}$Ar,
this excited configuration is essentially equivalent to the 4p--4h superdeformed 
configuration in $^{36}$Ar.

Cranking is then performed to study high-spin states. 
In the proton sector, the behaviors of single-particle orbitals are similar to 
those of $^{36}$Ar \cite{HFB}. 
For example, the occupation numbers in the $\pi$p$_{3/2}$ orbital monotonically
decrease up to $J=16 \hbar$ while the $\pi$f$_{7/2}$ orbital starts to increase 
at $J=$ 8 $\hbar$ due to the disappearance of the pairing gap energy.

In the neutron sector, clear differences are observed from the $^{36}$Ar case.
The occupation number in the $\nu$f$_{7/2}$ orbital is almost constant ($\sim$3) 
against the total angular momentum; it is about 1.5 times larger than that for 
$^{36}$Ar. 
The $\nu$d$_{5/2}$ orbitals are almost fully occupied ($\simeq 5.5$) 
from the low- to the high-spin regions.
In the case of $^{36}$Ar, the structural change involving the sharp backbending
is caused by a particle feeding from the p$_{3/2}$ orbital to the f$_{7/2}$ 
orbital for both protons and neutrons.
In the neutron sector of $^{40}$Ar, this feeding happens from the p$_{3/2}$
to the many other single-particle orbitals.
This is because the rise of the neutron Fermi level enhances the occupation 
numbers of the single-particle orbitals, particularly at the bottom end of 
the {\it fp} shell. 
For example, the f$_{7/2}$ is well occupied by $\simeq 40\%$. 
This high occupation influences the response of the system to the Coriolis 
force. 
In general, low-$\Omega$ states tend to come down energetically lower, so that 
such states are the first to be "submerged" when the Fermi level rises. 
As a result, neutron states near the Fermi level in $^{40}$Ar possess a higher 
$\Omega$ value and the rotational alignment is suppressed. 
In $^{36}$Ar, many $\Omega = 1/2$ states are vacant in the {\it fp} shell, 
so that it is possible to place particles in the $\Omega = 1/2$ states during 
the feeding from the p$_{3/2}$ orbital to the f$_{7/2}$ orbital.
However, in $^{40}$Ar, such $\Omega = 1/2$ states are filled due to the
rise of the neutron Fermi level. 
It is thus necessary to place neutrons in the $\Omega$ = 3/2 or 5/2 levels 
in order to generate angular momentum. 
But this way of feeding weakens the rotational alignment.
This ``Pauli blocking effect'' is one of the reasons why $^{40}$Ar does not 
backbend (at least, not in the spin region so far observed). 
It is also worth mentioning that because of the rise of the neutron Fermi level 
in $^{40}$Ar, angular momentum generation is spread among the extra f$_{7/2}$ 
neutrons, in comparison with $^{36}$Ar. 
This means that, unlike their neutron counterparts, the f$_{7/2}$ protons do not 
need to "work hard" to generate angular momentum. 
As a result, simultaneous alignment of the f$_{7/2}$ protons and neutrons 
does not occur in $^{40}$Ar. 
Our calculation confirms this picture.

In summary, a discrete-line superdeformed band has been identified in $^{40}$Ar. 
The observed large transition quadrupole moment ($Q_{t}= 1.45_{-0.31}^{+0.49} e$b)
supports its SD character. 
The properties of the SD band could be reasonably well explained by cranked HFB 
calculations with the P+QQ force. 
This finding of the SD band in $^{40}$Ar is similar to those observed in $^{36}$Ar 
\cite{ar36} and $^{40}$Ca \cite{ca40}, indicating the persistence of the SD shell 
structure in the neutron-rich A $=$ 30 $\sim$ 40 nuclei and possibly implying that 
$^{40}$Ar is a doubly SD magic nucleus with Z $=$ 18 and N $=$ 22.
The observed SD structure with a deformation of $\beta_2 \sim$ 0.5 caused by the 
mp--mh excitation across the {\it sd--fp} shell gap might explain
the origin of the strongly deformed ground state in the 'island of inversion'.

The authors thank the staff at the JAEA tandem accelerator for providing 
the $^{18}$O beam.

%
%
\begin{table}[H]
\caption{\label{occupation}Occupation numbers of $^{40}$Ar at $J = 0~\hbar$
determined by HFB calculation. Holes are denoted by negative numbers.}
\begin{center}
\begin{tabular}{lcccccccc}
\hline
Orbital & d$_{5/2}$ & s$_{1/2}$ & d$_{3/2}$ & f$_{7/2}$ & p$_{3/2}$ & f$_{5/2}$ & p$_{1/2}$ & g$_{9/2}$ \\
\hline
Proton  & -1.19     & -1.05     & -2.74     & 1.53      & 0.70      & 0.15      & 0.13      & 0.46 \\
Neutron & -0.65     & -0.39     & -2.02     & 2.88      & 0.95      & 0.44      & 0.27      & 0.53 \\
\hline
\end{tabular}
\end{center}
\end{table}

%
%
\begin{figure}[H]
\centering
\resizebox{13cm}{!}{
\includegraphics{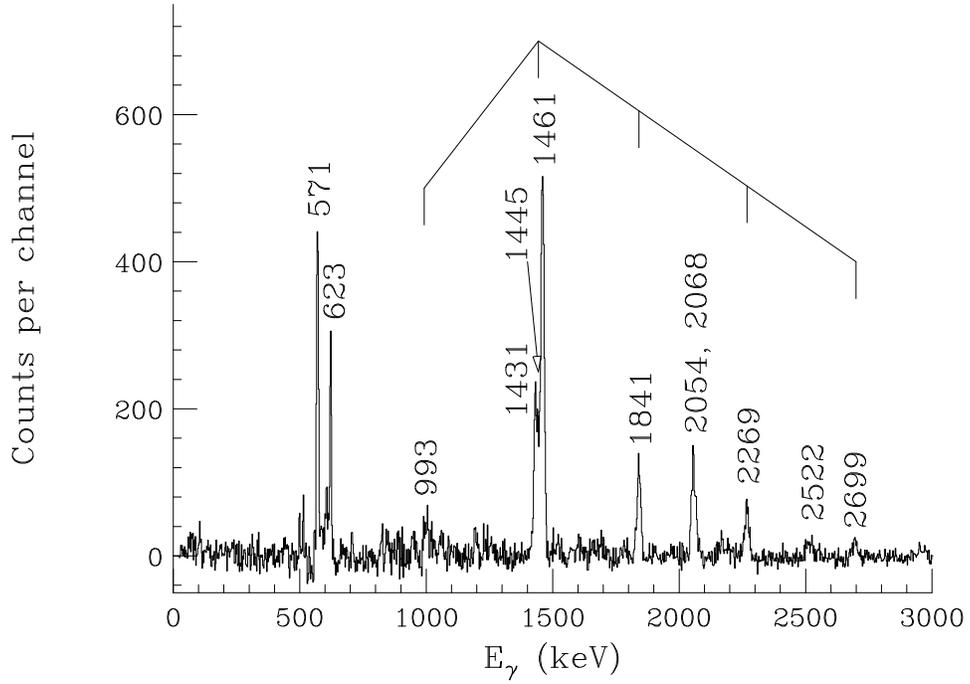}
}
\caption{\label{fig1}
Gamma-ray energy spectrum created by gating on in-band transition of
the SD band in $^{40}$Ar.
}
\end{figure}

%
%
\begin{figure}[h]
\centering
\resizebox{12cm}{!}{
\includegraphics{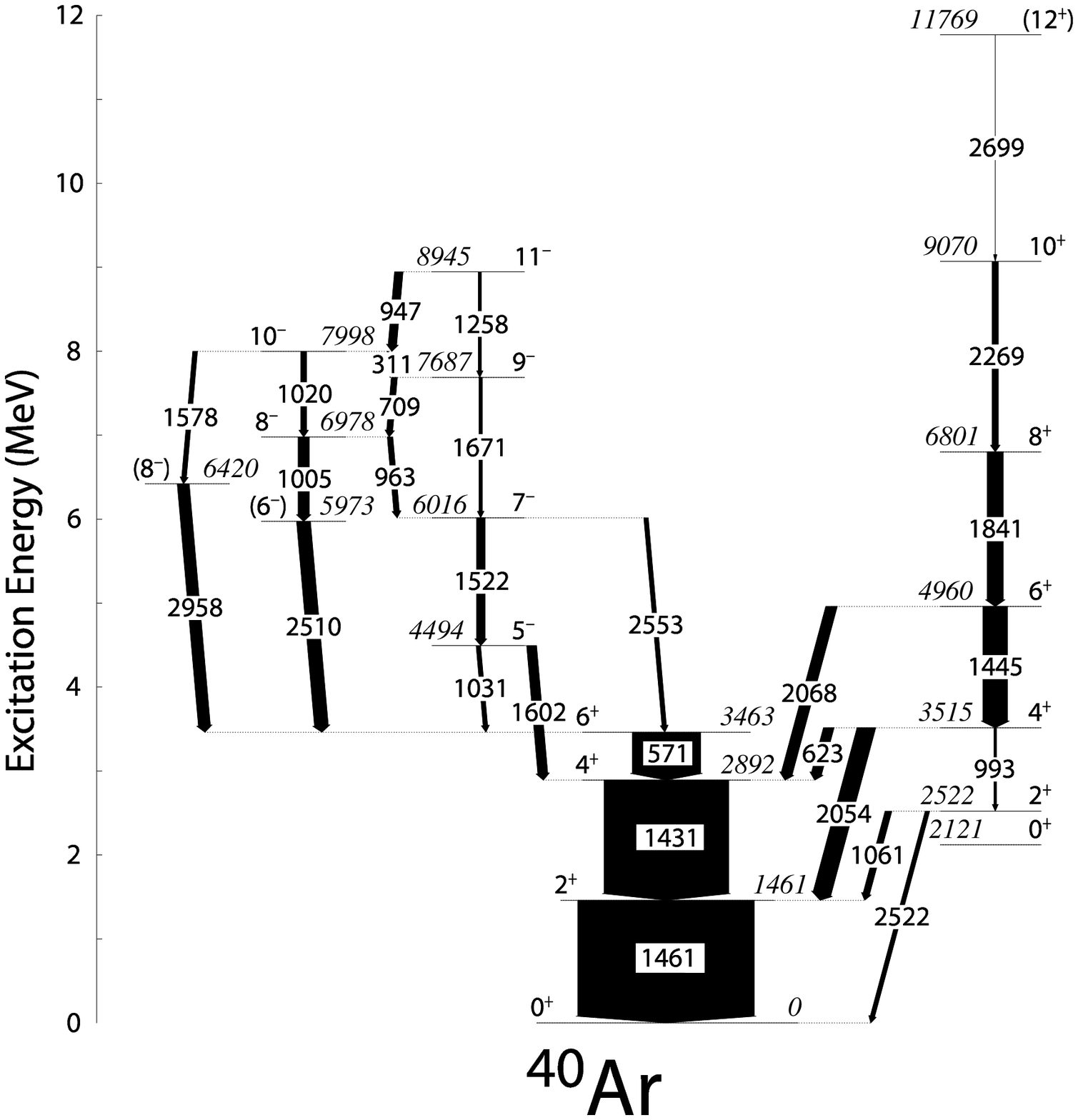}
}
\caption{\label{fig2}
Partial level scheme of $^{40}$Ar constructed in the present study.
The width of the arrow of each transition is proportional to its intensity.
}
\end{figure}

%
%
\begin{figure}[H]
\centering
\resizebox{12cm}{!}{
\includegraphics{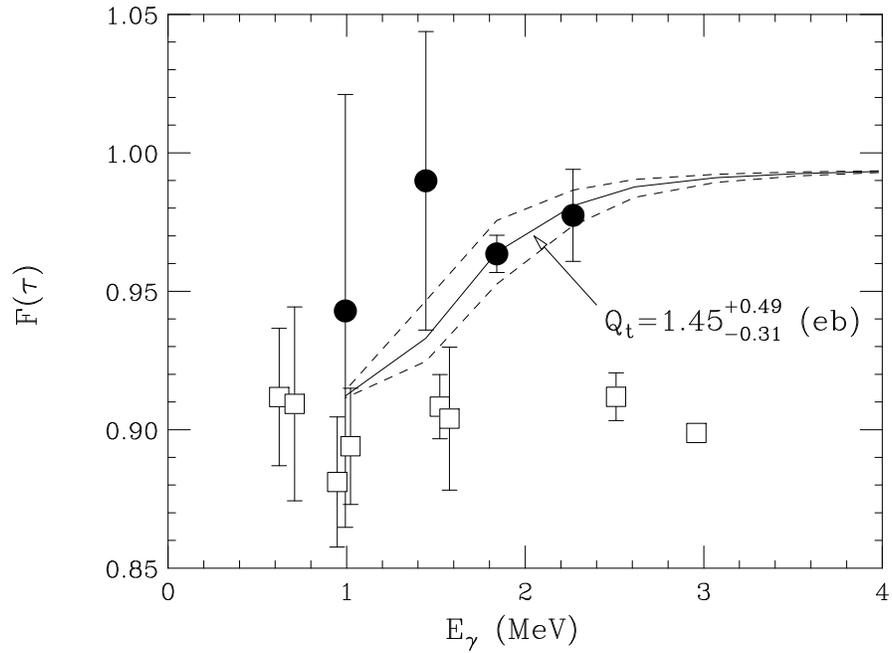}
}
\caption{\label{fig3}
Fractional Doppler shift F($\tau$) as a function of $\gamma$-ray energy.
Data points for the SD band (filled circles) and other transitions with
similar spins (open squares) are extracted from the residual Doppler shift
of the $\gamma$-ray energies. Solid lines represent the transition quadrupole
moment $Q_{t} = 1.45 eb$ and dashed lines correspond to the quoted 
uncertainties.
}
\end{figure}

%
%
\begin{figure}[H]
\centering
\resizebox{12cm}{!}{
\includegraphics{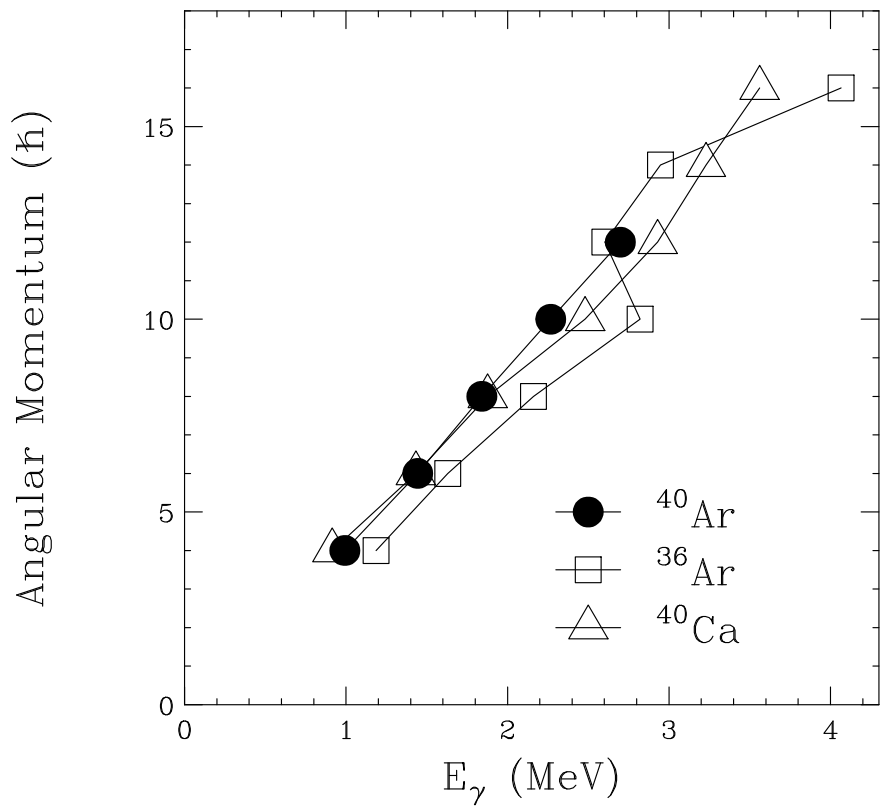}
}
\vspace{-3mm}
\caption{\label{fig4}
Backbending plot of the SD bands in $^{36}$Ar, $^{40}$Ar and $^{40}$Ca.
}
\end{figure}

\end{document}